\documentclass[english]{iopart}

\usepackage[T1]{fontenc}
\usepackage[latin1]{inputenc}
\usepackage{babel}
\usepackage{amsfonts}
\usepackage{amssymb}
\usepackage{graphics}
\usepackage{graphicx}
\usepackage[T1]{fontenc}
\usepackage[latin1]{inputenc}
\usepackage{epsf}

\begin{document}

\title[\Large The Fuchsian  equation of the 2-D Ising 
 $\chi^{(3)}$  susceptibility ]{\Large The Fuchsian  differential equation 
of the square lattice Ising model
 $\chi^{(3)}$  susceptibility }
\author{ 
N. Zenine$^\S$, S. Boukraa$^\dag$, S. Hassani$^\S$ and
J.-M. Maillard$^\ddag$}
\address{\S  C.R.N.A.,
Bld Frantz Fanon, BP 399, 16000 Alger, Algeria}
\address{\dag Universit\'e de Blida, Institut d'A{\'e}ronautique,
 Blida, Algeria}
\address{\ddag\ LPTL, Universit\'e de Paris 6, Tour 24,
 4\`eme \'etage, case 121, 4 Place
Jussieu, 75252 Paris Cedex 05, France} 
\ead{maillard@lpthe.jussieu.fr, maillard@lptl.jussieu.fr sboukraa@wissal.dz, njzenine@yahoo.com}

\date{\today}

\vskip .1cm
\begin{abstract}
Using an expansion method in the
variables $\, x_i$ that appear
in the $(n-1)$-dimensional integrals representing 
the $n$-particle contribution
to the Ising square lattice model
susceptibility $\chi$, 
we generate a long series of coefficients for
the 3-particle contribution
$\chi^{(3)}$, using a $\, N^4$ 
polynomial time algorithm. We give the Fuchsian differential equation of
order seven for $\chi^{(3)}$ that 
reproduces all the terms of our long series. An analysis of the properties 
of this Fuchsian differential 
equation is performed.
\end{abstract}
\vskip .1cm

\noindent 

\noindent \ams{ 34M55, 47E05, 81Qxx, 32G34, 34Lxx, 34Mxx, 14Kxx }

\vskip .1cm
\begin{indented}
\item[]{Keywords:  Susceptibility of the Ising model, series expansions,
Fuchsian differential equations, holonomy theory, apparent 
singularities, indicial equations, rigid local systems.}
\end{indented}

\section{Introduction}
\label{Intro}

Since the work of T.T. Wu {\it et al.} \cite{wu-mc-tr-ba-76}, 
it is known that the expansion in $n$-particle contributions to the zero field 
susceptibility of the square lattice Ising model
at temperature $T$ can be written as a sum:
\begin{eqnarray}
\label{coywu}
\chi(T) = \sum_{n=1}^{\infty} \chi^{(n)}(T) 
\end{eqnarray}
of $(n-1)$-dimensional 
integrals~\cite{nappi-78,pal-tra-81,yamada-84,yamada-85,nickel-99,nickel-00},
the sum being restricted to odd (respectively even) $n$ for the high
(respectively low) temperature case.
While the first contribution in the sum, $\chi^{(1)}$, 
is obtained directly without
integration, and the second one, $\chi^{(2)}$,
 is given in terms of elliptic integrals,
{\em no closed forms} for the higher order contributions are known
despite the well-defined forms of these $(n-1)$-dimensional integrals.
The $\, \chi^{(n)}$'s are $(n-1)$-dimensional 
integrals of holonomic (algebraic) expressions, and are consequently
{\em holonomic, or  ``D-finite''}: 
 they are solutions of {\em finite order} differential 
equations with polynomial coefficients ~\cite{lipsh,zilb}. Unfortunately,
 such theorems
of holonomy theory {\em do not give any lower or upper bound, or hint},  
as to the order of the differential 
equation satisfied by a given $\chi^{(n)}$. 

As far as  singular points are concerned (physical or 
non-physical singularities in the complex plane), 
and besides the known $\, s= \pm 1$ and $\, s\, = \pm i$ singularities, 
B. Nickel showed~\cite{nickel-99} 
that  $\, \chi^{(2\, n+1)}$ is singular\footnote{The 
singularities being logarithmic branch points of order 
$\, \epsilon^{2\, n\, (n+1)-1} \cdot ln(\epsilon)$
with $\, \epsilon \, = \, 1-s/s_i$ where $\, s_i$ is one 
of the solutions of (\ref{sols}).
} for the following finite values of
$\, s=sh(2\,J/kT) \,$ lying on the {\em unit circle} ($m=k=0$ excluded):
 \begin{eqnarray}
\label{sols}
&&2 \cdot \Bigl(s \, + \, {{1} \over {s}}\Bigr) \, = \, \, \, \, 
u^k\, + \,{{1} \over {u^k}} \,
+ \, u^m\, + \,{{1} \over {u^m}} \,\nonumber \\
&&  \qquad u^{2\, n+1} \, = \, \, 1,  \qquad   \qquad  
-n \,  \,\le\,  \, m, \, \, k\,  \,\,\le \, \, n  
\end{eqnarray}
When $n$ increases, the singularities of the higher-particle components 
of $\chi(s)$ {\em accumulate on the unit circle} $\, |s|\, = \, 1$.
The existence of such a natural boundary for the total $\,\chi(s)$, 
shows that  $\,\chi(s)$ is {\em  not D-finite} (holonomic) 
as a function of $\, s$.
To understand the analytical structure
of such a {\em transcendental} function, it is thus crucial 
to better understand the analytical structure of the
 ``holonomic'' $\chi^{(n)}$'s,
and, as a first step, to find the still unknown differential 
equation verified by the 3-particle contribution $\chi^{(3)}$. 

A significant amount of work had already been performed to
generate isotropic series coefficients
for $\chi^{(n)}$ (by B. Nickel \cite{nickel-99,nickel-00} up to order 116,
then to order 257 by 
A.J. Guttmann\footnote{Private communication.} {\it et al.}).
More recently,  W. Orrick {\it et al.}~\cite{or-ni-gu-pe-01b},
have generated coefficients\footnote{The short-distance terms  were shown to
have the form $\, (T-T_c)^p \cdot (log|T-T_c|)^q\, $
with $\, p \ge q^2$. }
 of $\chi (s)$ up to order 323 and 646 for high and low temperature 
series in $s$,
using  Perk's  {\em non-linear Painlev\'e difference equations}
for the correlation
 functions~\cite{or-ni-gu-pe-01b,coy-wu-80,perk-80,jim-miw-80,or-ni-gu-pe-01}.
As a consequence of this  non-linear Painlev\'e difference equation
and the associated remarkable recursion on the coefficients, the computer 
algorithm had a $\, O(N^6)$ polynomial growth 
of the calculation of the series expansion 
instead of an exponential growth
that one would expect
at first sight. However, in such a non-linear, non-holonomic,
Painlev\'e-oriented approach,
 one obtains results directly for the total susceptibility $\chi (s)$
 which does not satisfy
any linear differential equation, and thus prevents the easily 
disentangling of the contributions of the various holonomic $\chi^{(n)}$'s. 

In contrast, we develop here, a {\em strictly holonomic approach}. 
This approach enabled us to get 490 coefficients 
of the series expansion of $\, \chi^{(3)}$, from which we have deduced
the {\em Fuchsian differential equation} of
order {\em seven} satisfied by $\chi^{(3)}$. 
This Fuchsian differential equation presents a large
set of remarkable properties and structures that are briefly
sketched here and will be analyzed in more details in 
forthcoming publications, along with
 the analytical behavior of the solutions.
 The method used in this paper to obtain the Fuchsian
differential equation is not  specific of the third contribution $\chi^{(3)}$
and can be generalized, mutatis mutandis, to 
the other $\chi^{(n)}, \; (n>3)$ without any 
drastic changes in the mathematical framework\footnote{But 
certainly requiring larger computer calculations.}.

\section{Generating the series for $\chi^{(3)}$}
\label{generat}
Similarly to Nickel's papers~\cite{nickel-99,nickel-00},
we start using the multiple integral form of the $\,\chi^{(n)}$'s.
We focus  on $\chi^{(3)}$
 and consider the double integral:
\begin{eqnarray}
\label{chi3tild}
\chi^{(3)} (s) &=& {{ (1-s^4)^{1/4} } \over {s}} 
 \;\; \tilde{\chi}^{(3)} (s) \nonumber \\ 
\tilde{\chi}^{(3)} (s) &=& {\frac{1}{4\pi^2}}\int_0^{2\pi} d\phi_1 
\int_0^{2\pi} d\phi_2 \; \;
\tilde{y}_1 \; \tilde{y}_2 \; \tilde{y}_3 \,  
\Bigl( {\frac{1+\tilde{x}_1\tilde{x}_2\tilde{x}_3}
{1-\tilde{x}_1\tilde{x}_2\tilde{x}_3}} \Bigr)
 \; H^{(3)} 
\end{eqnarray}
with:
\begin{eqnarray}
\label{varxy}
\tilde{x}_j&=&{\frac{s}{1+s^2-s\cos{\phi_j}
+\sqrt{(1+s^2-s\cos{\phi_j})^2-s^2}}},  \\
\tilde{y}_j&=&{\frac{s}{\sqrt{(1+s^2-s\cos{\phi_j})^2-s^2}}}, \quad\quad
j=1,2,3, \quad \phi_1+\phi_2+ \phi_3=0 \nonumber
\end{eqnarray}
Many forms \cite{nickel-99,nickel-00} for $H^{(3)}$ may be taken and are
equivalent for integration purposes, e.g.,
\begin{eqnarray}
\label{h3old}
H^{(3)}=f_{23} \; \Bigl( f_{31}+ {{f_{23}}\over{2}} \Bigr) , \quad \quad
f_{ij}&=&(\sin{\phi_i}-\sin{\phi_j}) \;
 {\frac{\tilde{x}_i\tilde{x}_j}{1-\tilde{x}_i\tilde{x}_j}} 
\end{eqnarray}
It is straightforward to see that $\, \tilde{\chi}^{(3)}$
is only a function of the variable 
$w=\, {{1}\over{2}}s/(1+s^2)$. 
From now on, we thus focus on $\, \tilde{\chi}^{(3)}$
seen as a function of the well-suited variable $\, w$ instead of $\,s$.
We may expand the integrand in (\ref{chi3tild}) in this variable 
$\,w$ and integrate
the angular part. For $\,\tilde{\chi}^{(3)}$, this would mean there
are  18 sums to carry out.
We, instead, expand the integrand in the variable $\,\tilde{x}$ 
of (\ref{varxy}),
where we succeeded in deriving remarkable
formulas for $\tilde{y}\,\tilde{x}^n$ carrying one summation index.
As a consequence, we are able to write $\,\tilde{\chi}^{(3)} (w)$
as a fully integrated expansion. Our algorithm  runs
in a {\em polynomial} time calculations (namely $\, O(N^4)$).
In contrast with Orrick {\it et al}. calculations~\cite{or-ni-gu-pe-01b},
this calculation is not based on any recursion: it allows one to obtain 
any given coefficient {\em separately without 
requiring the storage of all the previous data}. 
The details of these calculations, tricks\footnote{For instance, 
our calculations also
underline the important role played by hypergeometric functions.}, 
and of this program will  be given elsewhere~\cite{ze-bo-ha-ma-04}. 

At present, we have obtained\footnote{We can get a new
coefficient every two days.
} the expansion of 
$\, \tilde{\chi}^{(3)}$ up to $\, w^{490}$. As expected,
this expansion is in agreement with the previous
results published by Nickel~\cite{nickel-99,nickel-00} 
using a numerical method of integration, 
as well as the improved unpublished results 
(up to $\, w^{257}$) by 
A.J. Guttmann~\footnote{Private communication.} {\it et al.} 
In terms of the well-suited variable $\, w$, the first terms
 of the expansion 
of $\, \tilde{\chi}^{(3)} (w) $ read :
\begin{eqnarray}
\label{chi3tildser}
&&{{\tilde{\chi}^{(3)} (w)} \over {8}}  \, = \, \, 
{w}^{9}+36\,{w}^{11}
+4\,{w}^{12}+884\,{w}^{13}+196\,{w}^{14} \nonumber \\
&& \qquad  \qquad \qquad +18532\,{w}^{15}+6084\,{w}^{16}\,+ \cdots 
\end{eqnarray}

\section{The Fuchsian differential equation satisfied by 
$\, \tilde{\chi}^{(3)} (w)$}
\label{thefuchs}
Given the expansion of $\, \tilde{\chi}^{(3)} (w) $
up to $\, w^{490}$, 
the next step will be to encode all the numbers in this long series 
into a linear differential equation. Note that such an 
equation should exist \cite{lipsh,zilb}
though, its order is unknown. 
Using a dedicated program for searching for such a linear differential equation
with polynomial coefficients in $w$ and steadily increasing the order, 
we succeeded finally in finding
the following linear differential equation of order 
{\em seven} satisfied by the 490 terms we have calculated for
$\, \tilde{\chi}^{(3)}$:
\begin{eqnarray}
\label{fuchs}
&& \sum_{n=0}^{7}\, a_n \cdot 
{{d^n } \over {dw^n}} F(w) \, \, = \, \, \, \, \, 0
\end{eqnarray}
with :
\begin{eqnarray}
\label{defQ}
&&a_7 \, = \, \, \nonumber \\
&& w^7 \cdot  \left(1- w \right) \,\left( 1 + 2\,w \right) \,
  {\left(1- 4\,w \right) }^5\,{\left( 1 + 4\,w \right) }^3\,
  \left( 1 + 3\,w + 4\,w^2 \right) \, P_7(w)  \nonumber \\
&&a_6 \, = \, \, w^6 \cdot {\left(1- 4\,w \right) }^4
\,{\left( 1 + 4\,w \right) }^2\, P_6(w)\nonumber \\
&&a_5 \, = \, \,
w^5\cdot {\left(1- 4\,w \right) }^3\,
\left( 1 + 4\,w \right) \,  P_5(w) \nonumber \\
&&a_4 \, = \, \, w^4 \cdot \left(1- 4\,w \right)^2 \,P_4(w), \qquad \quad \quad
a_3 \, = \, \, w^3 \cdot \left(1-4\,w \right) \,P_3(w)\nonumber \\
&&a_2 \, = \, \, w^2\,P_2(w) , \qquad \quad
a_1 \, = \, \, w\, P_1(w) ,\qquad \quad  a_0 \, = \, \, P_0(w)
\end{eqnarray}
where $P_7(w), P_6(w)$ $\cdots$, $P_0(w)$
 are polynomials of degree respectively
28, 34, 36, 38, 39, 40, 40 and 36 in $\, w$~\cite{ze-bo-ha-ma-04}. These 
polynomials are
too large to be given here.

Note that the series of $\tilde{\chi}^{(3)} (w)$ up to order 359,
 is sufficient to identify the
differential equation when the order $q=7$
and the successive degrees $47, 46, 45, 44, 43, 42, 41$ and $36$ 
of the polynomials in front of the $F^{(n)}(w)$'s are imposed.
Inside this framework, and since our series has 
490 coefficients, we have here
$\, 131\,$ verifications of the correctness 
of this differential equation.

At this point, let us remark that if one had worked with 
variable $s$, instead of $w$,
the series in $s$ up to order 699 would be needed in order to obtain
 the seventh order differential equation
satisfied by $\tilde{\chi}^{(3)} (s)$.

Since the singular points of this differential equation
 correspond to the roots 
of the polynomial corresponding to the highest
 order derivative  $\,  F^{(7)}(w)$,
namely $a_7$,
we give the exact expression of $\, P_7(w)$ :
\begin{eqnarray}
\label{defP7}
&&P_7(w) \,\, =\,\, 
    1568 + 15638\,w - 565286\,w^2 - 276893\,w^3  \nonumber \\
&& \qquad  +   34839063\,w^4   + 100696470\,w^5  - 1203580072\,w^6  \nonumber \\
&& \qquad   - 5514282112\,w^7 + 18005067728\,w^8 + 110343422816\,w^9 \nonumber \\
&& \qquad -140604884224\,w^{10} - 1825536178688\,w^{11}   + 920432273408\,w^{12} 
 \nonumber \\
&&  \qquad + 28913052344320\,w^{13} +    38181758402560\,w^{14} 
 \nonumber \\
&&  \qquad- 112307319603200\,w^{15} -544140071665664\,w^{16}
 \nonumber \\
&& \qquad
 - 1144172108054528\,w^{17}- 1027222993371136\,w^{18} \nonumber \\
&&  \qquad -1992177026596864\,w^{19} -
    2948885085421568\,w^{20} \nonumber \\
&& \qquad  + 2211524294737920\,w^{21}  +8204389336481792\,w^{22} \nonumber \\
&& \qquad  + 675795924156416\,w^{23} - 
    2882636020187136\,w^{24} \nonumber \\
&&  \qquad - 5364860829302784\,w^{25} -
    222238787764224\,w^{26} \nonumber \\
&&  \qquad + 158329674399744\,w^{27}    +39582418599936\,w^{28} 
\end{eqnarray}
The differential equation (\ref{fuchs}) is an equation of the {\em Fuchsian
type since there are no singular points, finite or infinite, other
 than regular singular points}. With this property, using 
Frobenius method~\cite{ince-56}, it is straightforward 
to obtain, from the indicial equation,
 the critical exponents, in $\, w$, for each regular singular point. These
 are given in Table 1.

\vskip .2cm 
\begin{tabular}{|l|l|l|}\hline
&& \\
$w=0 	$&$			s=0$&$ 				\rho=9, 3, 2, 2, 1, 1, 1$ \\
&& \\
$w=-1/4$&$			s=-1 $&$				\rho=3, 2, 1, 0, 0, 0, -1/2$ \\
&& \\
$w=1/4$&$				s=1$&$				\rho=1, 0, 0, 0, -1, -1, -3/2$ \\
&& \\
$w=-1/2$&$			1+s+s^2=0$&$			\rho=5, 4, 3, 3, 2, 1, 0$ \\
&& \\
$w=1$&$				2-s+2s^2=0	$&$		\rho=5, 4, 3, 3, 2, 1, 0$ \\ 
&& \\
$1+3\, w+4\,w^2 =0$&$		(2s^2+s+1)(s^2+s+2)=0$&$	\rho=5, 4, 3, 2, 1, 1, 0$ \\ 
&& \\
$1/w=0$&$				1+s^2=0	$&$		\rho=3, 2, 1, 1, 1, 0, 0$ \\
&& \\
$w=w_P$, 28 roots&	$s=s_P$, 56 roots &$		\rho=7, 5, 4, 3, 2, 1, 0$ \\
&& \\
  \hline 
\end{tabular}
\vskip .2cm 
{\bf Table 1:} Critical exponents for each regular singular point.
$w_P$ is any of the 28 roots of $P_7(w)$. We have also shown the
corresponding roots in the $s$ variable\footnote{In the variable $\, 
s$ the local exponents for $w=\pm 1/4$
are twice those given.}.
\vskip .4cm 

At this point, it is worth recalling the Fuchsian relation on 
 Fuchsian type equations. Denoting
 $\, w_1$,  $\, w_2,\,  \cdots \, $,  
$\, w_m$, $\, w_{m+1} \, = \, \infty$, the regular 
singular points of a  Fuchsian type equation
of order $\, q$
and $\, \rho_{j,1}, \, $ $\, \cdots, \, \rho_{j,q}$
  ($j \, = \, 1, \, \cdots\, , m+1$)
the $\, q$ roots of the indicial equation corresponding to each 
regular singular point $\, w_j$, the following  Fuchsian relation 
holds :
\begin{eqnarray}
\label{Fuchs}
\sum_{j\, =\, 1}^{m+1} \,\sum_{k\, =\, 1}^{q} \, \rho_{j,k}\,\,  = \, \,\, 
{{ (m-1)\, \, q \, \, (q-1)} \over {2}}
\end{eqnarray}
The number of regular singular points is here $\, m+1 \, = \, 36$
corresponding respectively to the 28 roots of $\, P_7$,  the two 
roots of $\, 1+3 w\, +4\, w^2$, five regular singular points,
and the point at infinity $\,w \, = \,\infty$.
The Fuchsian relation is actually verified here with $\, q\, = 7$,
$\, m\, = \, 35\,$, the sum of all the  $\, \rho_{j,k}$'s 
being actually $\, 714 \, $.

Let us comment on the singularities appearing in (\ref{fuchs}).
The well-known ferromagnetic and antiferromagnetic
critical points
correspond respectively to
 $\, w =1/4$ and $\, w=-1/4$, while  $\, w =0$
corresponds to the zero or infinite temperature points.
The value $\, w = \infty$, or $\, s\, =\,\pm i $, 
is known as a non-physical singularity. The values 
    $\, w =\,1\,$ and $\, w=\,-1/2$ correspond to the
non-physical singularities (\ref{sols2})
 described by Nickel for $\, n=1\, $ ($m=k=0$ excluded) :
 \begin{eqnarray}
\label{sols2}
&&{{1} \over {w }} \, = \, \, \, \, 
u^k\, + \,{{1} \over {u^k}} \,
+ \, u^m\, + \,{{1} \over {u^m}} \,\nonumber \\
&&  \qquad u^{3} \, = \, \, 1,  \qquad   \qquad  
-1 \,  \,\le\,  \, m, \, \, k\,  \,\,\le \, \, 1  
\end{eqnarray}

Furthermore, besides the known singularities mentioned above, we remark
the occurrence of the roots of the polynomial $\, P_7$ of degree 28 in $w$, and
the two {\em quadratic numbers}  $\,1 +3\, w +4\,w^2 \, = \,0$
which are not of the form (\ref{sols2}). The two quadratic numbers
are not on the $s$-unit circle : $|s|=\, \sqrt{2}$ and  $|s|=\, 1/\sqrt{2}$.

To analyze the local solutions of the differential equation, 
let us recall that, in general, it is known~\cite{ince-56} that for a set 
of $k$ local exponents such that the difference, in absolute
value, between any two of them is an integer, the local
 solutions may contain logarithmic
terms up to $\log^{k-1}$. In fact, the Fuchsian 
equation (\ref{fuchs}) is such that
the occurrence of logarithmic terms, in the local solutions near a given
regular singular point, is due only to the occurrence of multiple roots in the
corresponding indicial equation: a root of
 multiplicity $p$ inducing logarithmic
terms up to $\log^{p-1}$, i.e., {\em at most}
 $\log^2$ for (\ref{fuchs}), as it is
shown also in the monodromy matrices (see below).

More precisely for $P_7$, near any of its roots, all the 
local solutions
carry {\em no logarithmic terms}  
and {\em are analytical} since the exponents {\em are 
all positive
integers}. The roots of $P_7$ are thus {\em apparent 
singularities}
\cite{ince-56} of the Fuchsian equation (\ref{fuchs}).
Note that the "apparent" character of $P_7$, 
means that, to
ensure the absence of logarithmic terms in the general 
solution,
there are $\, q \cdot (q-1)/2=21$ relations between the various 
$P_n$'s at each root
of $P_7$ (for more details see \cite{ince-56}).

The two unexpected quadratic numbers correspond to
singularities of solutions of (\ref{fuchs}) which behave 
locally like
$(1+3 w\, +4\, w^2) \cdot \ln(1+3 w\, +4\, w^2) \cdot q_1
\, + \, q_2$,
where $\, q_1$ and $\, q_2$ are analytic functions near 
the two quadratic roots 
$\, 1+3 w\, +4\, w^2\, = \, 0$.
This ``weakly'' singular behavior can also be seen on
the monodromy matrix (see below) associated with these 
two roots,
where one finds a nilpotent matrix of order 
two (no $\, log^2\, $ term).

Details on the local solutions of (\ref{fuchs})
around each regular singular point,
together with the constants giving the particular physical solution
$\tilde{\chi}^{(3)}$, will be given in a forthcoming publication.

We will sketch here the analysis of the Fuchsian equation by
focusing on the local solutions of (\ref{fuchs}) around  $\, w \,= \, 0$.
Recall that the exponents  near $\, w \, = \, \, 0$ 
are respectively: $\, 9, \, 3, \, \, 2\, , 1$ and again 
$2\, , 1, \, 1$, all of them integers. 
Besides the $\tilde{\chi}^{(3)}/8$
solution of the Fuchsian differential equation 
which behaves, near $\, w \, = \, \, 0$, like (\ref{chi3tildser}), 
(which we will denote, from now on, $\, S_9$, since its leading term
behaves like $\, w^9$), 
one expects
solutions with leading terms behaving like $\, w \, $, 
 $\, w^2 \, $ and  $\, w^3$, because of the previous integer exponents
 near $\, w \, = \, \, 0$. Furthermore, 
because of the singularity confluence (repetition of same exponents)
and the occurrence of integer exponents, one also expects 
 solutions with the above-mentioned logarithmic terms.

Actually, we have found
two  remarkable {\em rational and algebraic} 
solutions of (\ref{fuchs}), namely :
\begin{eqnarray}
\label{remarkable}
S_1 \, = \, \, {{w} \over {1\, -4\, w}}, \qquad \qquad 
S_2 \, = \, \,
{\frac {{w}^{2}}{ \left( 1-4\,w \right) \sqrt {1-16\,{w}^{2}}}}
\end{eqnarray}
a solution  behaving like $\, w^3$,
that we denote $\, S_3$ :
\begin{eqnarray}
&&S_3 \, = \, \, 
{w}^{3}+3\,{w}^{4}+22\,{w}^{5}+74\,{w}^{6}+417\,{w}^{7}+1465\,{w}^{8}
\nonumber \\
&& \qquad \qquad +7479\,{w}^{9}  +26839\,{w}^{10}\, + \cdots 
\end{eqnarray}
and three solutions  $\, S_1^{(2)}$, $\, S_2^{(2)}$ and $\,S_1^{(3)}$ 
with logarithmic terms, behaving, at small $\, w$, as:
\begin{eqnarray}
\label{other}
S_1^{(2)}\, &=& \, \, \,  w \cdot \ln(w) \cdot c_1\, -32\, w^4 \cdot c_2 \\
\label{other2}
S_2^{(2)}\, &=& \, \, \,  
\, w^2 \, \ln(w) \cdot d_1\, + \, 8\, w^4 \cdot d_2  \\
\label{other3}
S_1^{(3)}\, &=& \, \, 3\, w \cdot \ln(w)^2 \cdot e_1\, 
-12 \, w^3 \cdot \ln(w) \cdot e_2\, -19\, w^4 \cdot e_3 
\end{eqnarray}
where $\, c_1, \, c_2, \, d_1, \, d_2, \, e_1, \, e_2, \,e_3$
denote functions that are
analytical at  $\, w\, = \, 0$. 
However, these functions are
not all independent. They can also be written in terms of 
$S_1,\, S_2,\, S_3$ and $S_9$ as follows :
\begin{eqnarray}
\label{others}
&&S_1^{(2)}\, = \, \, \, \ln(w) \cdot
(S_1-4\, S_2\, +16\, S_3\, -216\, S_9)\, -32\, w^4 \cdot c_2 \\
&&S_2^{(2)}\, = \, \, \,  
\ln(w) \cdot 
(S_2\, -2\, S_3\, +24\, S_9)\, + \, 8\, w^4 \cdot d_2 \nonumber \\
&&S_1^{(3)}\, = \, \, 3 \, \ln(w)^2 \cdot (S_1\, +5\, S_2\, -2\, S_3)\, \,  \nonumber \\
&&\qquad \qquad \quad - 6 \,  \, \ln(w) 
\cdot (2 \, S_3 \, -S_1^{(2)} \,-9\, S_2^{(2)} )
\,\,  -19\, w^4 \cdot e_3 \nonumber
\end{eqnarray}
where :
\begin{eqnarray}
\label{other4}
 c_2 \, &=& \, \, 
1 \, +{\frac {167}{96}}\,w+{\frac {2273}{96}}\,{w}^{2}
+{\frac {6977}{120}}
\,{w}^{3}+{\frac {19371}{40}}\,{w}^{4}+\cdots \nonumber \\
 d_2 \, &=& \, \,
1+{\frac {5}{2}}\,w+{\frac {103}{4}}\,{w}^{2}
+{\frac {315}{4}}\,{w}^{3}+{\frac {2191}{4}}\,{w}^{4}+\cdots
\nonumber \\
 e_3 \, &=& \, \,
1+{\frac {7693}{456}}\,w
+{\frac {575593}{11400}}\,{w}^{2}
+{\frac {2561473}{5700}}\,{w}^{3}
+{\frac {127434803}{93100}}\,{w}^{4}+\cdots
\nonumber 
\end{eqnarray}
Denoting $\, \Omega \, = \, 2\, i\, \pi$,
one immediately deduces
from the previous relations (\ref{others}),
the monodromy around $\, w \,= \, 0$ :
\begin{eqnarray}
&&S_1^{(2)} \, \,\rightarrow \, \qquad \, S_1^{(2)} \, \,
+\, \Omega  \cdot (S_1-4\, S_2\, +16\, S_3\, -216\, S_9)
\nonumber \\
&&S_2^{(2)} \, \,\rightarrow \,\qquad \, S_2^{(2)}
+\, \Omega  \cdot (S_2-2\, S_3\, +24 \, S_9)
\nonumber \\
&&S_1^{(3)} \, \,\rightarrow \,\qquad \, S_1^{(3)}
-6\, \Omega  \cdot  ( 2\, S_3\, - S_1^{(2)} \, -9\,S_2^{(2)}) \nonumber \\
&& \qquad \qquad \quad \qquad + 3\, \Omega^2 \cdot (S_1\, +5\,S_2\, -2S_3)
\nonumber
\end{eqnarray}
This calculation is a straight consequence
of the fact that one only has a ``logarithmic''
monodromy, which just amounts to changing
$\, \ln(w)$ into $\, \, \ln(w)\, + \, \Omega$, 
in the previous expressions.
Denoting $\, Id_7$
the  $\, 7 \times 7$ identity matrix, and 
taking the following order $\, S_1$, $S_2$, $S_3$, $S_9$, 
$\, S_1^{(2)} $, $\, S_2^{(2)} $, $\, S_1^{(3)}$ 
for our seven-dimensional basis of solutions of (\ref{fuchs}),
one can see that any $\, n$-th power of the 
$\, 7 \times 7$ monodromy matrix $\, M$, $\, M_n \, = \, M^n$, 
satisfies $\,M_n^3 \,- 3\, M_n^2  +3\, M_n \, -Id_7 \, = \, \, 0$, and can
 be written as 
the following sum of the identity matrix, 
and the two first powers of the order-three nilpotent matrix $\, N$
($N^3 \, = \, 0$) :
\begin{eqnarray}
\label{Mn}
&&M^n  \, \, = \, \, \, \, \, \, 
Id_7\, + \, n \cdot \Omega \cdot N \, 
+ \,{{ (n \cdot \Omega \cdot N)^2} \over {2}} \, \, = \, \, \,
\exp(n \cdot \Omega \cdot N),  \\
&&\nonumber \\
&&\qquad  \mbox{where :} \qquad \quad \quad 
 N \, = \, \, 
\left[ \begin {array}{ccccccc}
 0&0&0&0&1&0&0\\
\noalign{\medskip}0&0&0&0&-4&1&0\\
\noalign{\medskip}0&0&0&0&16&-2&-12\\
\noalign{\medskip}0&0&0&0&-216&24&0\\
\noalign{\medskip}0&0&0&0&0&0&6\\
\noalign{\medskip}0&0&0&0&0&0&54\\
\noalign{\medskip}0&0&0&0&0&0&0
\end {array} \right]
\nonumber
\end{eqnarray}
\vskip .2cm 

As it should, one can check that the form (\ref{Mn}), valid 
for any positive {\em or negative} integer $\, n$, is such that :
$\, M^n \cdot M^m \, = \, M^{m+n}$. 

The monodromy matrices corresponding to the other
regular singular points yield very similar 
results.
Those corresponding to $\, w \, = \, \pm 1/4$ are
matrices of determinant  $\, - 1$, while all the others are
of determinant  $\, + 1$.
 The monodromy around 
the 28 roots of $\, P_7$ is of course trivial (apparent singularities). The 
monodromy matrices corresponding to the other
regular singular points verify very simple relations like the previous
relation 
$\,M_n^3 \,- 3\, M_n^2  +3\, M_n \, -Id_7 \, = \, \, 0 $
or $\,M_n^2 \, - 2\, M_n \, +Id_7 \, = \, \, 0 $, and also
$\,M_n^6 \,- 3\, M_n^4  +3\, M_n^2 \, -Id_7 \, = \, \, 0 $ for those
 corresponding to $\, w \, = \, \pm 1/4$.
This is a straight consequence of the fact that, in a Jordan form,
 only very simple Jordan blocks occur\footnote[1]{We
 thank Jacques-Arthur Weil for
this result.}, such as :
\begin{eqnarray}
\label{Jordan}
J_2 \, = \, \, \left[ \begin {array}{cc} 1&1\\
\noalign{\medskip}0&1\end {array} \right], 
\qquad \qquad 
J_3\, = \, \, \left[ \begin {array}{ccc}
 1&1&0\\
\noalign{\medskip}0&1&1\\
\noalign{\medskip}0
&0&1\end {array} \right]
\end{eqnarray}
For instance, the monodromy matrices 
around the points $\, w \, = \, 0$ and $\, w \, =\, \infty$
are on the same footing, both requiring 
one  $\, J_3$ Jordan block, one $\, J_2$ block, and
 the $\, 2 \times 2$ identity matrix. There 
is  another set of 
monodromy matrices corresponding to the points  $\,w \, = \, \, 1$,
$\, w \, = \, \, -1/2$ and the two roots of 
$\, 1\, +3\,w \, + 4\, w^2\, = \, 0$
(associated with $\, J_2$ blocks). More precisely, 
the monodromy matrices for $\, w\, = \, 0$ and $\, w \, = \, \infty$
verify $\,M_n^3 \,- 3\, M_n^2  +3\, M_n \, -Id_7 \, = \, \, 0 $,
but for $\, w\, = \, -1/2$ and $\, w \, = \, 1$
they verify $\,M_n^2 \, - 2\, M_n \, + Id_7 \, = \, \, 0 $, 
as well as for the 
two roots of $\, 1+3\, w\, + 4 \, w^2\, = \, 0$.

A detailed analysis of the monodromy group will be performed
elsewhere. However, such an analysis of the Galois group\footnote[2]{The main difficulty
is to find a {\em global} structure like the monodromy Galois group from
the knowledge of all these {\em  local} monodromy matrices expressed
in the different 
well-suited {\em local} basis associated with each regular singular point.} becomes easier
to perform when taking into account some remarkable factorization 
and decomposition properties that are sketched in the next
section. 

\section{Algebraic properties of the Fuchsian equation}
\label{sketch7}
Let us define :
\begin{eqnarray}
\label{lambda}
&&\lambda \, = \,  \nonumber \\
&&{w}^{2}\,  
\left( 1+3\,w+4\,{w}^{2} \right)^{5} \,
\left( 1+2\,w \right)^{3} \,
\left(-1+4\,w \right)^{47/2} \,
 \left( 1+4\,w \right)^{31/2} \,
\left(1-w \right)^{3} \nonumber \\
&& f \, =\, (1-w) \,(1+2w)\,(1-4w)^5 \,(1+4w)^3 \,(1+3w+4w^2) \nonumber
\end{eqnarray}
We denote $\, L_7$ the seventh order linear differential operator,
corresponding to the Fuchsian differential equation (\ref{fuchs}):
\begin{eqnarray}
L_7 \, =\, {{d^7} \over{dw^7}}\,+\, {\frac{1}{a_7(L_7)}} \cdot
\sum_{k=0}^{6} a_k(L_7) \, {{d^k} \over{dw^k}}
\end{eqnarray}
where the $\, a_k(L_7)$'s are the polynomials $a_k$ defined in (\ref{defQ}).
Polynomial $a_7(L_7)$ reads 
$\, a_7(L_7)=\, w^7\cdot f \cdot P_7$, where we distinguish the
``actual'' and the apparent singularities.

We now give some remarkable factorization properties of the 
linear differential operator $L_7$.
The fact that the very simple expression $\, S_1$ (see (\ref{remarkable})) 
is a solution of the differential operator $\, L_7$ and
 that it is also solution of the
first order differential operator $\, L_1$ :
\begin{eqnarray}
\label{L1}
 L_1  \, = \, \, {{d} \over {dw}} \, -\, {\frac{1}{ w\, (1-4\, w)}}
\end{eqnarray}
implies the following factorization of $L_7$ 
(or more precisely the right-division of $L_7$ by $L_1$):
\begin{eqnarray}
\label{L7M6L1}
L_7 \, = \, \, M_6 \cdot  L_1
\end{eqnarray}
Similarly, the fact that $\, S_2$ is a solution of $\, L_7$, and
 that it is also clearly solution of a
first order differential operator $\, N_1$ :
\begin{eqnarray}
 N_1 \, = \, \,{{d} \over {dw}} \, -\,
 {\frac{2\, \left( 1+2\,w \right)}{w \left( 1-16\,w^2 \right)}}
\end{eqnarray}
implies the existence of another factorization of $\, L_7$ :
\begin{eqnarray}
L_7 \, = \, \, \, N_6 \cdot  N_1 
\end{eqnarray}
The linear differential operators of order 6 read ($X_6$ denoting $M_6$ or
$N_6$) :
\begin{eqnarray}
X_6 \, =\, \, {{d^6} \over{dw^6}}\,+\, {\frac{1}{a_6(X_6)}} \cdot
\sum_{k=0}^{5} a_k(X_6) \, {{d^k} \over{dw^k}}
\end{eqnarray}
where $a_k(X_6)$ are polynomials in $w$ and
$\, \, a_6(M_6) \, =\, a_6(N_6) \,=\, w^4 \cdot f \cdot P_7 (w)$.

These two factorizations are consequences of the existence 
of remarkable simple algebraic solutions of $L_7$. 
Besides this, there exists another factorization 
of $L_7$ related to the adjoint differential equation of (3.1). 
This is explained as follows.
One can also see that the adjoint of $L_7$,  denoted $L_7^*$, admits
the following rational solution\footnote[3]{We thank Jacques-Arthur Weil
for the remarkable result (\ref{S1star}) and (\ref{JAW}).}:
\begin{eqnarray}
\label{S1star}
S_1^{*} \, = \, \, \,
 {{f \cdot  Q_{6} (w) }\over { w^{3} \cdot P_7}} 
\end{eqnarray}
with :
\begin{eqnarray}
&&Q_{6} \, = \,\, 
1+19\,w-368\,{w}^{2}-3296\,{w}^{3}+17882\,{w}^{4}+272599\,{w}^{5}\nonumber \\
&&\qquad + 160900\,{w}^{6} -6979208\,{w}^{7}+7550800\,{w}^{8}+203094872\,{w}^{9}\nonumber \\
&&\qquad -278920192\,{w}^{10} -3959814304\,{w}^{11}-2115447424\,{w}^{12}
\nonumber \\
&&\qquad+20894729472\,{w}^{13} +39719728128\,{w}^{14}+20516098048\,{w}^{15}\nonumber \\
&&\qquad+256763363328\,{w}^{16}-327065010176\,{w}^{17}  -8810227761152\,{w}^{18}  \nonumber \\
&&\qquad 
+414933057536\,{w}^{19} +116411936538624\,{w}^{20}
\nonumber \\
&&\qquad +296827723186176\,{w}^{21}
+317648030138368\,{w}^{22}\nonumber \\
&&\qquad +179148186189824\,{w}^{23} +194933533179904\,{w}^{24}\nonumber \\
&&\qquad +112931870081024\,{w}^{25} -55246164328448\,{w}^{26}\nonumber \\
&&\qquad +11063835754496\,{w}^{27}+1511828488192\,{w}^{28}
\end{eqnarray}
\vskip .1cm 
yielding immediately the following factorization 
(or more precisely the left-division of $L_7$ by $M_1$):
\begin{eqnarray}
\label{JAW}
 L_7 \,  \,\,= \, \,\, \, 
M_1 \cdot  L_6, \, \qquad  \quad 
\mbox{where :}  \quad 
\quad  L_6 \, = \, \, L_5 \cdot N_1 
\end{eqnarray}
\begin{eqnarray}
\label{defM1}
\hbox{with:} \quad \quad\quad  \quad M_1 \, = \, \, \,  {{d} \over {dw}}
 + \,  {\frac{1}{S_1^*}} \, {\frac{dS_1^*}{dw}}
\end{eqnarray}
and where $\, L_5$ and  $\, L_6$  are 
linear differential  operators of order five and 
 six respectively, which read :
\begin{eqnarray}
\label{L6}
L_q \, =\, {{d^q} \over{dw^q}}\,+\, {\frac{1}{a_q(L_q)}} \cdot
\sum_{k=0}^{q-1} a_k(L_q) \, {{d^k} \over{dw^k}}, \qquad q=5,6
\end{eqnarray}
with :
\begin{eqnarray}
a_6(L_6) \, &=&\, w^6 \cdot  f  \cdot Q_{6}(w)  \nonumber \\
a_5(L_6) \, &=&\, w^5 \cdot  (1-4\,w)^4 \, (1+4\,w)^2   \cdot Q_{5}(w)  \\
a_5(L_5) \, &=&\, 
w^3 \cdot {\left(1- 4\,w \right) }\,{\left( 1 + 4\,w \right) }^2
\cdot f \cdot   Q_{6}(w) \nonumber
\end{eqnarray}
The roots of polynomial $Q_6 (w)$ are apparent singularities of the 
differential equations associated to the operators 
$L_6$ and $L_5$.

One can also see that $\, M_6$ in (\ref{L7M6L1}) 
can also be decomposed as follows :
\begin{eqnarray}
\label{M6M5T1}
&& M_6 \, = \, \, \, M_5 \cdot T_1 \qquad \qquad \mbox{with:} \quad \quad \nonumber \\
&& \qquad  M_5 \, =\, {{d^5} \over{dw^5}}\,+\, {\frac{1}{a_5(M_5)}} \cdot
\sum_{k=0}^{4} a_k(M_5) \, {{d^k} \over{dw^k}} \nonumber
\end{eqnarray}
where  $\; \; a_5(M_5)\, = \, \,
 w^4 \, {\left(1- 4\,w \right) }^5 \, {\left( 1 + 4\,w \right) }^4
\cdot f \cdot  P_7(w)$,  $ \, \, \, $ and:
\begin{eqnarray}
T_1 \, = \,\,{{d} \over {dw}}\, -\,
{\frac{1+4\,w+48\,w^2}{w\,  \left( 1-16\,w^2 \right)}}
 \nonumber 
\end{eqnarray}
Calculations seem to show that $\, L_5$ (resp. $\, M_5$) is 
irreducible : we have not found any 
further factorizations like
$\, L_5\, = \, L_4 \cdot A_1$ or $\, L_5\, = \, B_1 \cdot L_4$, or 
even $\, L_5\, = \, L_3 \cdot A_2$ or $\, L_5\, = \, B_2 \cdot L_3$.

One can also be interested in the Wronskians of $\, L_7$ and $\, L_6$. They 
can be written as follows  :
\begin{eqnarray}
\label{wronski}
W(L_7) \, \, = \, \, \, {{P_7} \over {\lambda}}, \qquad \qquad 
W(L_6) \, \, = \, \, \, {{f \cdot Q_{6} (w) } \over {  w^3 \cdot \lambda}}
\end{eqnarray}
thus checking the following relation 
on the Wronskians of $\, L_7$, $\,L_6$, and $\,M_1$, deduced from (\ref{JAW}), namely:
\begin{eqnarray}
\label{wronsknew3}
W(L_7) \, = \, \, W(M_1) \cdot  W(L_6)
\end{eqnarray}

From these factorizations, and/or decompositions,
relations, corresponding to the existence of {\em rational, 
or algebraic} (square root of rational), expressions, 
one should not be surprised to find that the 
Wronskians of all the various operators we define here, are remarkable
rational, or algebraic (square root of rational), 
expressions (see (\ref{wronski})). The fact that 
the squares of the Wronskians (\ref{wronski})
are rational functions corresponds to the following identities between
 $\, P_7$ and $\, P_6$ (resp. $\, Q_{6}$ and $\, Q_{5}$) :
\begin{eqnarray}
\label{wronsk21}
&& {{P_7'} \over {P_7}} \, - \, \, {{\lambda'} \over {\lambda}} 
+ {{(1-4\,w)^4 \, (1+4\,w)^2}\over{w \, f}} \cdot {{P_6} \over {P_7}} \, = \, 0 \\
\label{wronsk212}
&&  {{Q_{6}'} \over {Q_{6}}} \, 
-\, \,  {{3} \over {w}} \, +\, \, {{f'} \over {f}}
- \, \, {{\lambda'} \over {\lambda}}
+ {{(1-4\,w)^4 \, (1+4\,w)^2}\over{w \, f}} \cdot {{Q_{5}} \over {Q_{6}}} \, = \, 0
\end{eqnarray}

The Wronskians (or their inverse for the adjoints) 
of all the differential 
operators $\, L_7$, $\, L_7^{*}$, $\, M_6$,$\, M_6^{*}$,
$\, M_5$, $\, M_5^{*}$ can be expressed as simple powers of $\, P_7$
divided by a simple expression similar to $\, \lambda$ in (\ref{lambda}).
Similarly, the Wronskians (or their inverse for the adjoints) 
of all the differential 
operators $\, L_6$, $\, L_6^{*}$, $\, L_5$, $\, L_5^{*}$
 can be expressed as simple powers of
 $\, Q_{6}$
divided by a simple $\lambda$-like expression. All these Wronskians
are thus {\em rational} ($\, L_5$, $\, L_5^{*}$, $\, M_5$, $\, M_5^{*}$)
or such that their {\em square is rational} 
($\, L_7$, $\, L_7^{*}$, $\, L_6$, $\, L_6^{*}$,
$\, M_6$, $\, M_6^{*}$). Of course, there are relations between these
Wronskians which are in agreement with the operator factorizations 
previously described (see (\ref{wronsknew3})).

Besides the ``fundamental''
singularities $\, w\, = \, 0, -1/2, \, \pm 1/4, \, 1$
and the ``unexpected'' quadratic numbers singularities solutions of 
$\, 1+3\,w+4\,{w}^{2} \, = \, 0$,
one could say that a large part of the ``arithmetic complexity'' of the
operator $\, L_7$ arises from
 the two polynomials of degree 
28 in $w$,  $\,P_7$ and $Q_{6}$. 

The occurrence of the {\em apparent singularities}, 
associated with the quite large polynomial $\, P_7$,
can be considered slightly unpleasant or disturbing : 
one would like to exchange the seventh order Fuchsian 
equation (\ref{fuchs}), for another differential equation
(or a differential system) where these ``spurious'' singularities
have disappeared. This is the so-called {\em desingularization 
problem} for differential equations~\cite{Tsai,Abra}. We have performed such 
``desingularization'' : the ``price'' to be paid is that 
one has no longer a seventh order differential equation, 
but an {\em eighth} order differential equation. The associated eighth order
differential operator reads:
\begin{eqnarray} 
\label{desingu} 
&&L_8 \, =\,\, {{d^8} \over{dw^8}}\,+\, {\frac{1}{a_8(L_8)}} \cdot 
\sum_{k=0}^{7} a_k(L_8) \, {{d^k} \over{dw^k}} 
\qquad \quad \mbox{with :} \\ 
&&a_8(L_8) \, =\, \, \nonumber \\
&&w^{n_1} \,
\left( 1-4\,w \right) ^{n_2} \,
\left( 1+4\,w \right)^{n_3} \,
\left( 1+2\,w \right)^{n_4} \,
\left(1-w \right)^{n_5} \,
\left( 1+3\,w+4\,{w}^{2} \right)^{n_6} 
\nonumber 
\end{eqnarray} 
$n_1, \cdots, n_6 \,$ being positive integers.
The other polynomials $\, \,  a_k(L_8) $, are quite large
and will not be given here.
Let us sketch 
this desingularization procedure. 
Recall the exponents associated with the roots of
 $\, P_7$ (see Table 1). They are $\, 0, 1, 2, 3, 4, 5, 7$;  one sees
that the exponent $\, 6$ is missing. Basically, the method
 amounts to building a differential equation
having all the solutions of (\ref{fuchs}), together with a solution\footnote[9]{
The solution $\, S \, =\, P_7^6$ is too naive, it puicks out the 
apparent singularities of $\, P_7$ but introduces new spurious apparent singularities. }
associated with this missing exponent $\, 6$, in order 
to ``fill the gap''
and make  the roots of $\, P_7$ ordinary points for the
homogenous differential equation associated with $\, L_8$. Note too that the
desingularized differential equation, or equivalently 
the operator $\, L_8$ are far from being unique. The various eighth order $\, L_8$  operators
are, by construction, of the form $\, L_8 \, = \, \, \tilde{L}_1 \cdot L_7$, 
where the first order operator  $\, \tilde{L}_1$ is quite involved\footnote[5]{
As far as the solutions of  $\, L_8$ are concerned, the differential operator 
 $\, L_8$  adds 
to the known solutions of $\, L_7$, a solution with a
 strong exponential behavior.}. 
From a desingularized form like (\ref{desingu})
one can also introduce a differential
 system\footnote[6]{See for instance the chapter 6 of~\cite{Marius}.} :
\begin{eqnarray}
\label{systdesingu}
&&\theta \cdot Y \, = \, \,
 A \cdot Y, \qquad \quad 
 Y \, = \, (y, \,\, \theta \cdot y, \, \,
 \theta^2 \cdot y, \,\, \theta^3 \cdot y, \, \,\cdots)  \\
&& \mbox{with :}\quad \quad \quad  A \, = \, \, \, {{ A_0} \over {w}} \, +{{ 
A_{1}} \over {w-1/4}} \, 
+{{ A_{2}} \over {w+1/4}} \,+ {{ A_{3}} \over {w+1/2}} \, \nonumber \\
&& \qquad \quad \quad \quad \qquad +\, 
 {{ A_{4}} \over {w-1}} \,+ \, {{ A_{5} \,w+A_{6}} \over {w^2+3/4\,w+1/4}} 
\,+\, P(w) \cdot Id_{8}
\nonumber   
 \end{eqnarray}
where $Id_{8}$ denotes the $8 \times 8$ identity matrix,
 $\, P(w)$ denotes a polynomial in $\, w$, 
and where the $\, A_n$ matrices
are very simple   $8 \times 8$  matrices,
simply related to the monodromy matrices, and where $\, \theta$ denotes
the ``well-suited'' derivation associated with the ``true'' singularities :
 \begin{eqnarray}
 \theta \, = \, \,  w \cdot \left(1- w \right) \,\left( 1 + 2\,w \right) \,
  \left(1- 4\,w \right) \,\left( 1 + 4\,w \right)\,
  \left( 1 + 3\,w + 4\,w^2 \right)    \cdot {{d} \over {dw}} \nonumber
 \end{eqnarray}
A form like (\ref{systdesingu}) is clearly much simpler, and more canonical : 
the calculations one has in mind (Galois group, rigidity index 
(see below)) should be much simpler to perform with this 
canonical system form (\ref{systdesingu}).
Details will be given elsewhere.

\section{Comments and Speculations}
\label{comments}

A better understanding of the total susceptibility
$\, \chi$ certainly requires an exhaustive knowledge
of the singularities of the 
 successive Fuchsian differential equations
 associated with the $n$-particle contributions $\,\chi^{(n)}$. 
Besides the apparent singularities associated with the roots of the
polynomial $\, P_7$, we have noted the occurrence of 
the two rather unexpected {\em quadratic numbers} solutions of
 $\,1 +3\, w +4\,w^2 \, = \,0$, 
which, also, {\em are not of Nickel's 
 form} (\ref{sols2}). The elliptic parameterization 
of the Onsager model is well-known. Recalling the 
exact expression of the modular invariant  $\, {\cal M}\, $
for the Ising model (see for instance~\cite{Louis}) :
\begin{eqnarray}
\label{modular}
{\cal M} \, =\,\, \,{\frac {1}{1728}}\,{\frac 
{ \left( 1  -16\,{w}^{2}+ 16\,{w}^{4}\right)^{3}}{{w}^{8} 
\left( 1\, -16\,{w}^{2}\right) }}
\end{eqnarray}
it can be seen that 
the two {\em quadratic numbers}
correspond to a {\em rational value} of the modular invariant: 
$\,{\cal M}  \, = \, \, -125/64$, while $\, w =\,1\,$
also corresponds to a rational value $\,{\cal M}  \, = \, \, -1/25920$, 
that $\, w =\, \pm 1/4, \, 0, \,\infty $
correspond to $\,{\cal M}  \, = \, \, \infty$, and that $\, w =\,-1/2\,$
corresponds to the rational value $\,{\cal M}  \, = \, \, 32/81$. 
Could this mean 
that the Fuchsian equation (\ref{fuchs})
 could be ``canonically'' 
associated with an elliptic curve ? One can, for instance, 
recall the period mappings and Picard-Fuchs 
equations (and beyond mirror symmetries ...) 
 associated with a family of 
elliptic curves~\cite{Doran,Doran2}. The
 Picard-Fuchs equation :
\begin{eqnarray}
\label{Picard}
144\, s \, (s-1)^2 \cdot \Bigl( s \cdot {{d^2\, f} \over {ds^2}}\, + \, \,
 {{d\, f} \over {ds}} \Bigr)
 + \, \,(31\, s\, -4) \cdot f \, = \, \, \, 0 
\end{eqnarray}
is associated with the family of elliptic curves :
\begin{eqnarray}
y^2 \, = \,\, 4\, x^3 \, +{{27} \over {1-s}} \cdot x \, 
+{{27\, s} \over {1-s}} 
\nonumber 
\end{eqnarray}

We now make a few comments on the physical solution $S_9$.
The Fuchsian equation is highly non-trivial and structured : from the previous
 factorizations, and/or decompositions,
one might imagine that $S_9 \, = \,\tilde{\chi}^{(3)}/8 \,$
is in fact a solution of a sixth order homogeneous differential equation, or even a 
fifth order homogeneous differential equation. This is not the case. Relation
(\ref{JAW}) means  that $\,S_9 \, = \, \tilde{\chi}^{(3)}/8  \,  $
can be seen as a linear combination of $\, S_1$ 
and of a solution of a sixth order linear homogeneous differential
 equation, $\, S(L_6)$, associated with the operator $\, L_6$. The solution  
 $\,S_9 \, = \, \,\tilde{\chi}^{(3)}/8  \,  $ is actually such a linear combination
$\, S_9 \, = \, \, \alpha \cdot S_1\, + \, \, S(L_6)$, 
$\, \alpha$  being different from zero\footnote[7]{The coefficient $\, \alpha$
characterizing the ``projection'' of $\, S_9$ on $\, S_1$, can
 easily be calculated writing 
$\, L_6(S_9\, - \alpha \cdot S_1) \, = \, \, 0 \,$. One finds $\, \alpha=1/24$.}. 
The three-particle contribution, $\,\tilde{\chi}^{(3)} $, 
is {\em thus a solution of the seventh order differential 
equation} (\ref{fuchs}) {\em and not of a homogeneous linear differential
of smaller order}.

Coming back to the analysis of the  Galois
 monodromy group~\cite{Singer,Bertrand,Mitschi}
of the Fuchsian equation (\ref{fuchs}), it is clear that
the previous factorizations, and/or decompositions,
impose severe reductions of the group.
From these various factorizations, or decompositions,
of the linear operator $\, L_7\, = \, M_1 \cdot L_6 \, = \, \, 
 M_1 \cdot L_5 \cdot N_1 $, one can deduce that 
the Galois monodromy group is 
isomorphic to $\, Gal( L_6)$, the 
 Galois monodromy group of $\, L_6$. Introducing $\, Gal( L_5)$, the 
 Galois monodromy group of $\, L_5$, one deduces from $\, L_6 \, = \, \, 
L_5 \cdot N_1 $, that the operator $\, N_1$ injects $\, Gal( L_5)$
in $\, Gal( L_6)$. This does not mean that $\, Gal( L_6)$ is isomorphic to
 $\, Gal( L_5)$, but that knowledge of $\, Gal( L_5)$
is required to describe $\, Gal( L_6)$. 
Recalling the rationality of the Wronskian of $\, L_5$, 
one deduces that  $\, Gal( L_5)$ is a subgroup of $\, SL(5,C)$ 
and not
 $\, GL(5,C)$ (the rationality of the  Wronskian means that 
all the monodromy matrices of $\, L_5$ have $\, +1$ determinants). 
The monodromy matrices of $\, L_6$ (as well as the one of 
 the Fuchsian equation (\ref{fuchs})) have $\, \pm 1$ determinants.
The Galois group $\, Gal( L_6)$ is, up to a $\, Z_2$-graduation,
a subgroup of $\, SL(6,C)$. Therefore  the Galois group
of the Fuchsian equation (\ref{fuchs}) is represented by $\, 6 \times 6$ 
matrices with $\, \pm 1$ determinants, most of its structure requiring
the analysis of a $\, Gal( L_5)$ subgroup of  $\, SL(5,C)$. 
Of course, one can also understand the  Galois group  of 
 the Fuchsian equation (\ref{fuchs}) from the analysis
of the  Galois group  of $\, M_5$, namely $\,  Gal( M_5)$
which is also subgroup of  $\, SL(5,C)$ (isomorphic
 to $\, Gal( L_5)$). It  seems that $\, L_5$ is not 
self-adjoint\footnote[8]{However it can be shown that $\, L_6$ 
is not self-adjoint (modulo conjugation by an operator). $\, L_6$ has 
$\, S_2$ as a solution. If it were
self-adjoint, $L_6^*$ would have a similar quadratic solution which is not the case.} 
 (modulo conjugation by an operator) and
irreducible (see Section 4). 
The two facts seem to rule out the existence
of a symplectic structure. 
A more detailed analysis of the Galois group
of the Fuchsian equation (\ref{fuchs}), that is, of the  Galois
 groups $\, Gal( L_5)$, and $\, Gal( L_6)$,
will be given in forthcoming publications. 

Let us, now, focus again on the ``physical'' solution 
of the Fuchsian equation (\ref{fuchs}),
$\,  S_9 \, = \,\tilde{\chi}^{(3)} /8 $, and on its successive derivatives 
$\, S_9'$,  $\, S_9''$,  $\, S_9^{(3)}, \cdots$ 
We have seen that, as far as the {\em linear} dependence of these 
expressions is concerned, we have a {\em seven-dimensional} vector space 
($S_9$ is {\em not} a solution of a homogeneous sixth order
 differential equation\footnote[1]{To be totally rigorous one should add that
the minimal operator of  $\,S_9 $ is a factor of  $\,L_7 $, and since
these factors are of order 1 or 6, and that $\,S_9 $ does not vanish on these
factors, it must vanish on  $\,L_7 $.}). 
However, to better understand the ``true nature'' of the susceptibility
$\, \chi$, one would like to characterize the ``degree of transcendence''
of $\, S_9 \, = \, \,\tilde{\chi}^{(3)}/8$, that is the minimal number
of successive derivatives of $\,  S_9$ satisfying  an algebraic 
{\em non-linear} relation\footnote[6]{For instance the 
Weierstrass ${\cal P}$ function 
verifies  the non-linear relation $\, \, {\cal P}'^2 = \, 
 4\,{\cal P}^3  + g_2\,{\cal P} +g_3$. }.
The  Galois monodromy group gives a valuable information about this 
 ``degree of transcendence''. Let us consider the orbit of 
$S_9$ under the Galois monodromy group: one gets an algebraic variety
dense in the subspace of solutions. The dimension of this
algebraic variety is actually this degree of transcendence.
This analysis, however, requires an exact knowledge
 of the Galois monodromy group. 

The susceptibility
$\, \chi$ has been shown to be a transcendental 
(non-holonomic, non $D$-finite) function: it cannot be solution
of a {\em linear} differential equation, but this does not mean that
it cannot be  solution
of a  differential equation.  
Along the previous ``non-linear'' line, one should emphasize that
the possibility 
that $\, \chi$ could be solution of some (Painlev\'e-like ?) 
{\em non-linear} differential equation, is not yet totally ruled out ! 

Within the ``linear'' Picard-Vessiot Galois monodromy group
 framework, one can also try to evaluate
 the ``index of rigidity~\cite{Reiter,Katz}''
of our Fuchsian differential equation, or the 
``index of rigidity'' of the operators $\, L_6$ or $\, L_5$. 
Roughly speaking, this index of rigidity
corresponds to the number of parameters that can be 
introduced to deform the linear differential equation, 
keeping the local monodromy matrices fixed\footnote[2]{See the notion
of {\em rigid local systems}~\cite{Katz,Darmon}. Basically it amounts
to calculating the dimensions of
the  centralizers of all the monodromy matrices.}. For instance,
the hypergeometric functions are totally rigid : as a consequence, 
``almost everything'' can be calculated on such functions\footnote[3]{In the case
of totally rigid systems, N. Katz has shown that the solutions have 
a {\em geometric interpretation} : they can be seen as periods 
of some algebraic varieties~\cite{Katz}. }.
The differential operators $\, L_7$, $\, L_6$ or $\, L_5$
have many integer exponents, and, thus,  their solutions
have logarithmic behaviors $\, (w-w_0)^n \cdot \bigl( ln(w-w_0) \bigr)^m$
around each regular singular point $\, w_0$. However,
 recalling  (\ref{other}), (\ref{other2})
 and (\ref{other3}), and, more generally,
considering the behavior near all the singular points,
 one sees that {\em only} $\, ln(w-w_0)$ and $\bigl( ln(w-w_0) \bigr)^2$ 
behaviors occur\footnote[7]{This can be seen directly on the very simple 
Jordan blocks for the monodromy matrices (see (\ref{Jordan})).
},
thus {\em downsizing} (see for instance 
(4.31) in Chapter (4) of~\cite{Marius}) 
the {\em index of rigidity} of these differential equations.
This is a strong indication 
 that the Fuchsian differential equation
(\ref{fuchs}), or the differential equations associated with 
$\, L_6$ or $\, L_5$, {\em are extremely rigid}. This can also be seen 
on the various monodromy matrices. The calculation
of the index of rigidity, or in other words, the calculation
of the ``small'' number of deformation parameters of (\ref{fuchs})
(resp. $\, L_6$, $\, L_5$) will be performed elsewhere. This 
remarkable rigidity is not a surprise, when recalling the 
well-known isomonodromy theory of the Ising model, and 
in particular the occurrence of Painlev\'e equations 
for the correlation functions of the Ising model. Does the
deformation theory of (\ref{fuchs}) in this 
``small'' number of deformation parameters actually yield 
 Painlev\'e-like  equations ? This is an open question.
From a more down-to-earth viewpoint, this rigidity 
can be seen as ``inherited'' from the ``total'' rigidity 
of  hypergeometric functions: our calculations
for generating the $\, w$-series of $\,\tilde{\chi}^{(3)}  \, $
are actually ``flooded'' by linear 
combinations (with binomial coefficients)
of  products of hypergeometric functions (see \cite{ze-bo-ha-ma-04} for more details).

\section{Conclusion}
The linear differential equation we have found for 
the three-particle contribution, $\, \chi^{(3)} (w)$,
to the susceptibility of the square lattice Ising model,
is a highly structured and remarkable Fuschsian differential equation.
We have sketched many of its remarkable properties and symmetries.
It is also worth 
recalling that the 28 roots of the polynomial $\, P_7$ {\em are 
apparent singularities}. We have been able to desingularise 
the Fuchsian equation (\ref{fuchs}),
 in order to
get rid of these apparent singularities related to the
quite involved polynomial $\, P_7$. A deeper analysis
 of this Fuchsian equation (monodromy group, 
critical behavior of the solutions around the various 
singular points, ... ) will be given in 
forthcoming publications.

All these results
can be generalized, {\em mutatis mutandis},
to deduce the  Fuchsian equations corresponding to the 
other $n$-particle contributions $\, \chi^{(n)}$'s. 
The building of a computer program
with a polynomial growth algorithm which can be generalized, 
mutatis mutandis,
for the other $\, \chi^{(n)}$'s,
was the key ingredient to get our Fuchsian equation. 
The ideas developed to create such polynomial growth programs
underline the role played by hypergeometric functions,
coming from a large number of remarkable identities on 
the underlying variables of the problem.
One may also think that quite complicated ``fusion-type'' relations 
on these hypergeometric functions can exist. 
This is crystal clear in the case of the 
2-particle contribution $\, \chi^{(2)}$ ~\cite{ze-bo-ha-ma-04}. 
This however remains to be done in full.

Clearly, beyond $\, \chi^{(3)}$, a global understanding of the structure 
of the hierarchy of all the $\, \chi^{(n)}$'s
could be contemplated. 
A better understanding of the total susceptibility
$\, \chi$ certainly requires an exhaustive knowledge
of the singularities of the 
 successive $n$-particle contributions $\,\chi^{(n)}$, or
 equivalently, of
 the corresponding successive Fuchsian equations. As far as
 analytical properties are 
concerned, we saw the occurrence
in (\ref{fuchs}) 
of  unexpected
quadratic numbers singularities, $\,1+3w+4w^2=0$
(these two singularities are not on the $s$-unit circle:
$\,|s|=\sqrt{2}, \, 1/\sqrt{2}$). Curiously these unexpected singularities
correspond to a {\em rational} value of the 
 {\em modular invariant}. Could this mean 
that the Fuchsian equations for
the successive $\, \chi^{(n)}$'s are ``canonically\footnote[5]{
See for instance the Picard-Fuchs 
equation (\ref{Picard}) associated with a family of 
elliptic curves~\cite{Doran,Doran2}.
}'' associated with an elliptic curve ? 
Should we rather understand this  hierarchy 
of Fuchsian equations in a linear (or non-linear) 
{\em monodromy deformation theory} framework ?

\vskip .3cm 

\ack
We thank Jacques-Arthur Weil for outstanding comments 
and calculations on our Fuchsian equation.
J-M.M  would like
to thank A. J. Guttmann and W. Orrick for extensive discussions 
in Melbourne three years ago 
and for a large number of e-mail exchanges more recently. We 
would like to thank
particularly W. Orrick for extensive calculations on our series. 
We also thank Fr\'ed\'eric Chyzak for many 
discussions and comments concerning
the algolib library, the gfun and Mgfun Holonomy, 
Mgfun, Ore-algebra packages 
and holonomy theory. We also thank  Daniel Bertrand and Henri Darmon
for illuminating comments and Emma Lockwood for comments on the manuscript.
(S. B) and (S. H) acknowledge partial support from PNR3-Algeria.

\vskip 3cm

\end{document}